# Temperature dependence of plastic scintillators

L. Peralta[ab]

[a] Departamento de Física da Faculdade de Ciências da Universidade de Lisboa, Lisboa, Portugal
[b] Laboratório de Instrumentação e Física Experimental de Partículas, Lisboa, Portugal

ABSTRACT: Plastic scintillator detectors have been studied as dosimeters, since they provide a cost-effective alternative to conventional ionization chambers. On the other hand, several articles have reported undesired response dependencies on beam energy and temperature, which enhances the necessity to determine appropriate correction factors. In this work, we studied the light yield temperature dependency of four plastic scintillators (BCF-10, BCF-60, BC-404 and RP-200A) and two clear fibers (BCF-98 and SK-80). Measurements were made using a 50 kVp X-ray beam to produce the scintillation and/or radioluminescence signal. The 0 to 40 ºC temperature range was scanned for each scintillator, and temperature coefficients obtained.

## Introduction

Plastic scintillator detectors (PSDs) have long been used as dosimeters [1] due to their good characteristics. For a time, it was accepted that their response to radiation was temperature independent [2]. In a paper published in 2009, Nowotny and Taubeck [3] presented evidence for temperature dependence of the response of some polystyrene based scintillators. Later, in 2013 two papers [4,5] presented measurements on the BCF-12 and BCF-60 (also polystyrene based) reporting non-vanishing temperature coefficients for both scintillators. The results for BCF-12 have been confirmed by Lee et al. in a work published in 2015 [6]. These works focused on the widely used BCF-12 and BCF-60 scintillators, being the information on other plastic scintillators scarce. Do scintillators based on other materials (acrylic - PMMA, Polyvinyltoluene - PVT, etc) exhibit the same temperature behavior? Is the effect restricted to scintillation light emission, or other types of emission like radioluminescence are also affected by temperature? In the present work four different scintillators and two different optical cables have been study for their light yield temperature dependence when irradiated with a 50 kVp X-ray beam.

## Material and methods

*Experimental setup*

PSDs were exposed to a 50 kVp X-ray beam produced by a Philips Oralix dental X-ray tube. An irradiation time of 3 s was chosen for each exposure. A Farmer ionization chamber PTW 30013 [7] connected to a PTW UNIDOS® E electrometer [8] was used to monitor the X-ray beam stability. The chamber was placed between the tube exit and the PSD, approximately 30 cm from the tube (figure 1). The readings from the chamber were used to correct the signal obtained with the PSD. This procedure ensures that variation in signal in the PSD are due to temperature changes and not to a variation of the beam flux.

Four PSDs were made of different scintillator materials: RP-200A, PMMA based; BC-404, PVT based; BCF-10 and BCF-60 polystyrene based. Some of the scintillators properties are presented in table 1. Scintillators were coupled to a 2.0 mm Super Eska™ SK-80 (Mitsubishi) PMMA optical cable [9] using optical grease (BC-630 Silicone Optical Grease, Saint-Gobain) [10].

The scintillator and optical cable were protected from daylight by a black polyester sleeve (3.2 mm in diameter heat-shrinkable type sleeve). The optical cable was read by an Hamamatsu R647P PMT [11]. The PMT was directly connected to a Standard Imaging MAX 4000 electrometer, integrating the current from the PMT.

Two optical cables were also studied. The BCF-98, polystyrene based clear fiber and the SK-80 fiber, also used as an optical cable when coupled to a scintillator. For the used X-ray beam (50 kVp), the light produced in these optical cables is due to radioluminescence [12]. Some properties of the optical cables are presented in table 2.

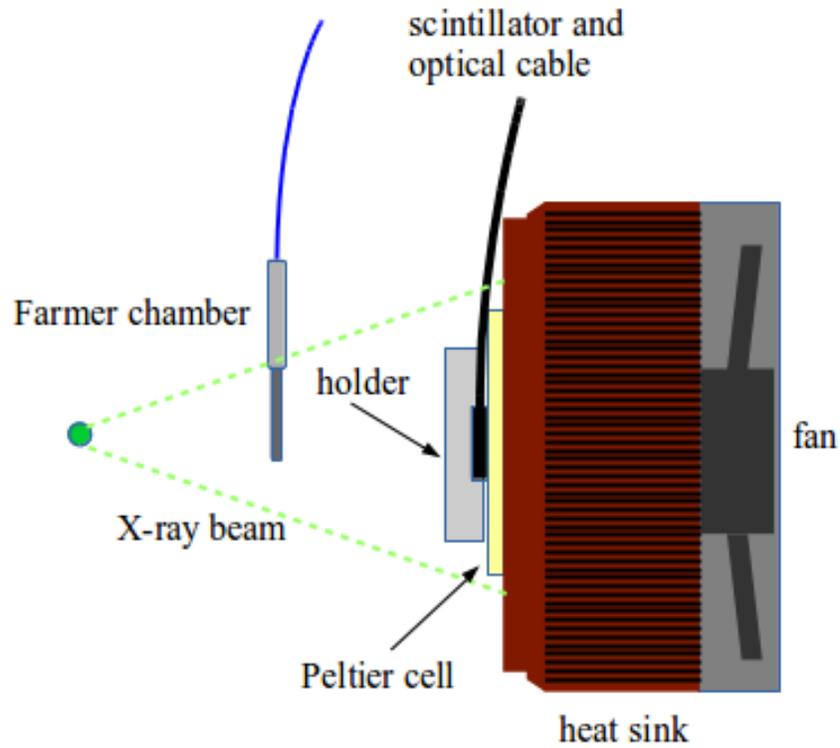

Figure 1. Artists view of the irradiation setup (not to scale).

Table 1. Scintillators' properties

| Scintillator | RP-200A | BC-404 | BCF-10 | BCF-60 |
| --- | --- | --- | --- | --- |
| Substrate material | PMMA | PVT | polystyrene | polystyrene |
| Manufacturer | Rexon | Saint-Gobain | Saint-Gobain | Saint-Gobain |
| Diameter (mm) | 3 | 4 | 2 | 2 |
| Length (mm) | 10 | 13.6 | 20 | 20 |
| Wavelength of max emission (nm) | 415 (1) | 408 (2) | 432 (2) | 530 (2) |
| Density (g/cm$^3$) | 1.18 (1) | 1.032 | 1.05 | 1.05 |

(1) Extracted from Rexon [13] datasheet, (2) Extracted from Saint-Gobain datasheet [14].

Table 2. Clear optical cable properties

| Cable | SK-80 | BCF-98 |
| --- | --- | --- |
| Substrate material | PMMA | polystyrene |
| Manufacturer | Mitsubishi | Saint-Gobain |
| Diameter (mm) | 2 | 2 |
| Irradiated length (mm) | 40 | 40 |
| Density (g/cm$^3$) | 1.19 | 1.05 |

The cooling system was constituted by a 40×40 cm$^2$ Peltier cell. The cell was also used as heating element, when polarity was reversed. Current was provided by a Tenman 72-10480 digital-control DC power supply capable of delivering up to 3 A.

*Temperature measurement*

Temperature was measured with a K-type thermocouple sensor connected to a multimeter (Benning MM 7-1). The measurement resolution is 0.1 ºC and the accuracy is ± 2 ºC [15]. The sensor was placed inside a black polyester sleeve, the same used to cover the scintillator and the optical cable. In this way, the sensor is subjected to the same thermal conditions as the scintillator. The sensor and scintillator were held in place, side by side, over the Peltier cell by a PMMA holder.

*Temperature dependence*

Ideally, the response from the dosimeter should be constant for a sufficiently wide range of temperatures. However, several studies now suggest that PSDs might have temperature dependency [3-6]. In order to correct the dosimeter readings, it is critical to study the scintillator's temperature dependence. In this study, we adopted a linear model [5] for the scintillator response over a limited range

$$S = S_0\big(1 + \alpha(T - T_0)\big) \quad (1)$$

where S is the scintillator response at temperature T, $S_0$ is the response at some reference temperature $T_0$ and α the temperature coefficient.

**Results and discussion**

The normalized response $S/S_0$ for the scintillators is presented in figure 2. As shown, the response variation with temperature is not linear and equation 1 can only be applied in a limited range. The reference temperature $T_0$ was chosen to be the nearest available value to 20 ºC. A linear fit is performed using data in the range 10 to 30 ºC to obtain the temperature coefficient. The same procedure was applied to the data obtained with the optical cables (figure 3). The obtained temperature coefficients are presented in table 3.

Except for BC-404, all the other scintillators present a clear temperature dependence within the measured range. The measured temperature coefficient value for BCF-10 is similar to the BCF-12 scintillator (α= (-2.6 ± 0.03) ×10$^{-3}$ ºC$^{-1}$) [6] (also a blue-emitting scintillator [14]). On the other hand, the value obtained for the BCF-60 scintillator is less than the values reported by earlier works [4,5] of about 5×10$^{-3}$ ºC$^{-1}$, but still showing a significant decrease on light yield with temperature. The PMMA based RP-200A scintillator also presented a measurable temperature dependence. Only the BC-404 PVT-based scintillator showed a negligible temperature dependence within the measured range.

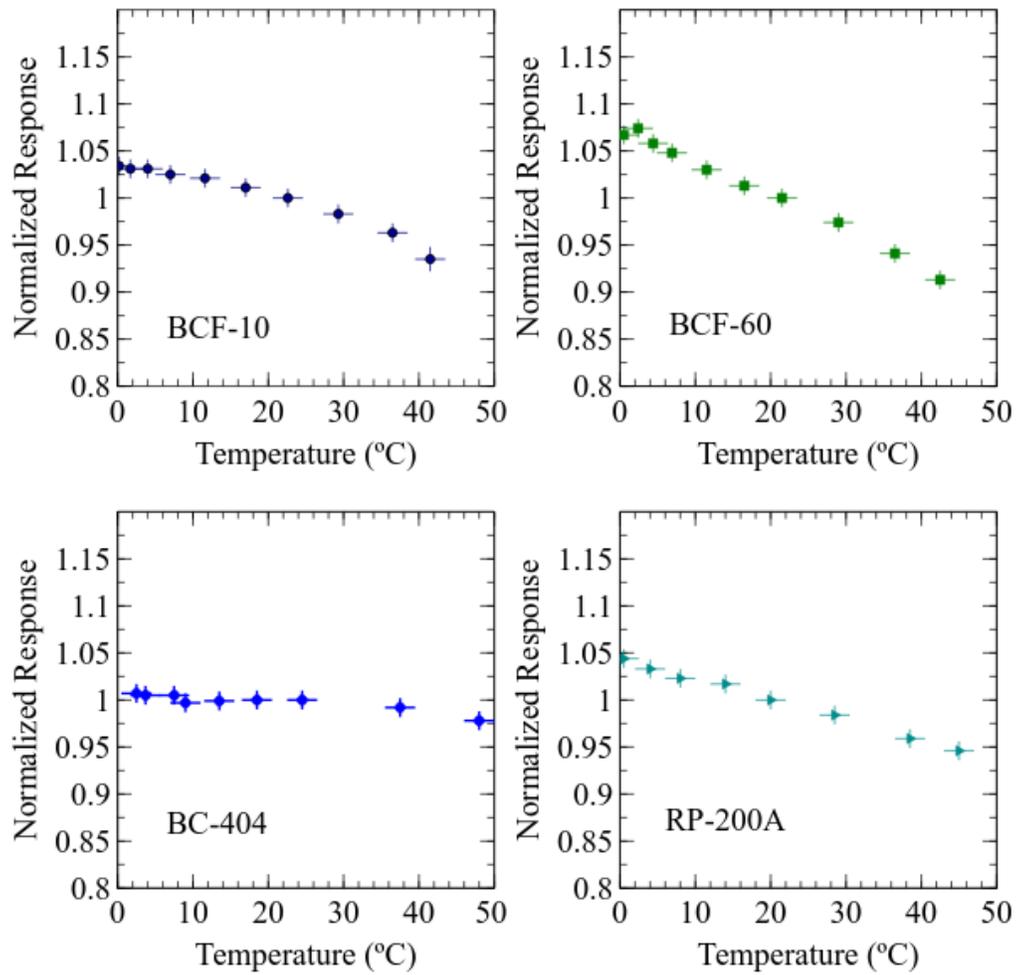

Figure 2. Normalized response of the scintillators as a function of temperature. The temperature $T_0$ was chosen to be the nearest value to 20 °C.

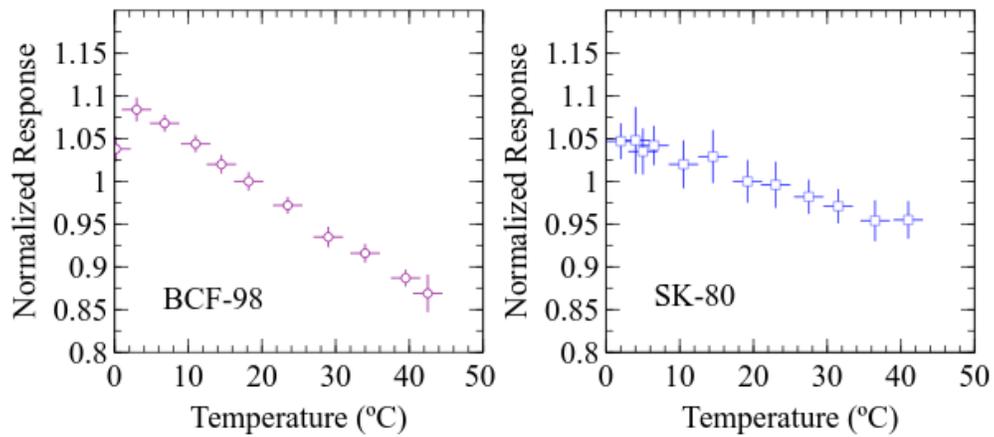

Figure 3. Normalized response of the optical cables as a function of temperature. The temperature $T_0$ was chosen to be the nearest value to 20 °C.

Table 3. Temperature coefficients obtained from the linear fit in the range 10 to 30 ºC. The presented uncertainties are one standard deviation.

| BCF-10 | BCF-60 | BC-404 | RP-200A | BCF-98 | SK-80 |
|---|---|---|---|---|---|
| $\alpha$ (ºC$^{-1}$) | $\alpha$ (ºC$^{-1}$) | $\alpha$ (ºC$^{-1}$) | $\alpha$ (ºC$^{-1}$) | $\alpha$ (ºC$^{-1}$) | $\alpha$ (ºC$^{-1}$) |
| $(-2.2 \pm 0.8) \times 10^{-3}$ | $(-3.2 \pm 0.8) \times 10^{-3}$ | $(0.08 \pm 1.3) \times 10^{-3}$ | $(-2.3 \pm 1.0) \times 10^{-3}$ | $(-6.1 \pm 0.8) \times 10^{-3}$ | $(-3.1 \pm 1.8) \times 10^{-3}$ |

On what concerned the optical cables, they display a temperature dependence, quite notorious for BCF-98. These results, in line with previous measurements by other authors [4-6], stress the need for more comprehensive studies on the subject in order to fully understand the underlying physical mechanisms of the light yield temperature dependence in plastic scintillators and plastic optical cables.

**Conclusion**

In this work temperature coefficients were measured for four scintillators and two optical cables. The obtained results are in line with previous studies for similar scintillators. The non-vanishing temperature coefficients confirm the small temperature dependence of the polystyrene and PMMA based scintillators. The effect is present in scintillators and optical cables. This is an indication the effect is already present at the substrate level. A smaller, nearly absent, temperature dependence was measured for the BC-404 PVT based scintillator. According to the manufacture datasheet [14] this is the expected behavior for the BC-4xx PVT based series in the -60 to 20 ºC range. The present work extends that range up to 50 ºC for the BC-404 scintillator. This scintillator has already been successfully used in PSDs prototypes tested in X-ray beam qualities used in tomosynthesis [16] and Cone Beam CT [17]. The temperature insensitivity of the BC-404 scintillator makes it a good candidate to be used in PSDs placed near the human body, where changes relative to room temperature can be expected.

**Acknowledgements**
We are grateful to Ashley Peralta for the review of the English text.**References**
[1] Sam Beddar and Luc Beaulieu (editors), Scintillation dosimetry, CRC Press, New York, 2016.
[2] A.S. Beddar, T. R. Mackie, and F. H. Attix, Water-equivalent plastic scintillation detectors for high-energy beam dosimetry: I. Physical characteristics and theoretical consideration, Phys. Med. Biol. 37 (1992), 1883–900.
[3] R. Nowotny and A. Taubeck, "A method for the production of composite scintillators for dosimetry in diagnostic radiology", Phys. Med. Biol. 54 (2009) 1457.
[4] L.Wootton and S. Beddar, "Temperature dependence of BCF plastic scintillation detectors", Phys. Med. Biol. 58 (2013) 2955–2967.
[5] S. Buranurak et al., "Temperature variations as a source of uncertainty in medical fiber-coupled organic plastic scintillator dosimetry", Radiation Measurements 56 (2013) 307-311.
[6] B. Lee et al., Effects of Temperature and X-rays on Plastic Scintillating Fiber and Infrared Optical Fiber, Sensors 15 (2015) 11012-11026, https://doi.org/10.3390/s150511012
[7] PTW 30013, http://www.ptw.de/farmer_chambers0.html, (accessed on August 2017)
[8] UNIDOS® E Universal Dosemeter, http://www.ptw.de/unidos_e_dosemeter_ad0.html, (accessed on August 2017)
[9] SK-80 fiber, http://i-fiberoptics.com/fiber-detail.php?id=130, (accessed on August 2017)
[10] Saint-Gobain 2005 Detector Assembly Materials, http://www.crystals.saint-gobain.com/products/assembly-materials, (accessed August 2017)
[11] R647P Phtotomultiplier Tube, http://www.hamamatsu.com/jp/en/R647P.html, (accessed on August 2017)
[12] A.R. Beierholm et al., A comparison of BCF-12 organic scintillators and Al2O3:C crystals for real-time medical dosimetry, Radiat. Meas. 43 (2008) 898-903.
[13] http://www.rexon.com (accessed on November 2017).
[14] Saint-Gobain, Organic scintillation material and assemblies,5